\documentstyle[aps,amsfonts,epsfig]{revtex}

\makeatletter
\global\@specialpagefalse
\def\@oddhead{\hfill QCMP Theory 00-11}
\let\@evenhead\@oddhead
\makeatother

\begin{document}

\title{Enhanced Bound State Formation in Two Dimensions\\ 
via Stripe-Like Hopping Anisotropies}

\author{Saurabh Basu and R. J. Gooding}
\address{Department of Physics, Queens University, \\ 
Kingston, Ontario K7L 3N6, Canada} 

\author{P.W. Leung}
\address{Department of Physics, Hong Kong University of Science and 
Technology,\\ Clear Water Bay, Hong Kong}

\date{\today}

\maketitle


\begin{abstract}

We have investigated two-electron bound state formation in a square 
two-dimensional $t-J-U$ model with hopping anisotropies for
zero electron density; these anisotropies are introduced to mimic the 
hopping energies similar to those expected in stripe-like arrangements 
of holes and spins found in various transition metal oxides. In this
report we provide 
analytical solutions to this problem, and thus demonstrate that bound-state
formation occurs at a critical exchange coupling, $J_c$, that decreases to
zero in the limit of extreme hopping anisotropy $t_y/t_x \rightarrow 0$. 
This result should be contrasted with $J_c/t~=~2$ for either 
a one-dimensional chain, or a two-dimensional plane with isotropic hopping.
Most importantly, this behaviour is found to be qualitatively similar
to that of two electrons on the two-leg ladder problem
in the limit of $t_{\rm interchain}/t_{\rm intrachain}~\rightarrow~0$. 
Using the latter result as guidance, we have evaluated the pair correlation
function, thus determining that the bound state corresponds to one electron
moving along one chain, with the second electron moving along the 
opposite chain, similar to two electrons confined to move along parallel,
neighbouring, metallic stripes. We emphasize that the above results are 
not restricted to the zero density limit --- we have completed an exact 
diagonalization study of two holes in a $12\times 2$ two-leg ladder 
described by the $t-J$ model and have found that the above-mentioned 
lowering of the binding energy with hopping anisotropy persists near 
half filling.

\end{abstract}


The experimental verification for the existence of stripes in
layered transition metal oxides, such as the high-$T_c$ superconductors, 
is now robust \cite {johnt}, and yet an important question
remains: Does the presence of such rivers of charge help, hinder,
or even possibly create, the pairing instability that leads to
superconductivity? Simple model calculations that might shed light
on this question are clearly of value, and in this paper we present
one such study for a model Hamiltonian that we have considered in an
attempt to mimic some of the physics of the stripe phases.

Previous work on the magnetic properties of the very weakly doped
cuprates \cite {cneto} modelled the observed experimental support
for stripe correlations \cite {borsa} using an effective Hamiltonian 
in which a (spatially) anisotropic exchange interaction was implemented 
to represent the stripe-induced magnetic energy scales. That is,
in the direction parallel to the stripes the full local $Cu-Cu$
exchange would be present, while perpendicular to the stripes
a reduced exchange would be encountered across such rivers of charge.
Renormalized Hamiltonians of a similar simplifying spirit were also
used in other studies of the doped cuprates \cite {aharony,doniach,skyrms}.

To this end, we have considered an effective Hamiltonian in which a 
carrier's ability to move along the river of charge is far larger than
its freedom to move between the rivers. This reduced mobility
reflects a carrier's reduced ability to move in an antiferromagnetically
correlated region, such as exists between the rivers of charge. 
The simplest theoretical modelling of such a situation corresponds to 
a Hamiltonian with local magnetic exchange that is the same in all 
directions, but with a hopping parameter which is very different
in one direction than the other. (Several studies of holes
moving on 2-leg ladders near half filling have considered 
a similar problem, but all but one have employed a limit that is different 
to that which we have investigated. To be particular, these authors 
have considered the hopping between the chains of the ladder to
be larger than that along each chain\cite{rice93,noack96,noack97,bose99}.
The recent work of Weihong, {\it et al.,}\cite{weihong} considers
a limit similar to that of this paper, although their focus is
two holes near half filling, not two electrons near zero filling.)

As a consequence of the above considerations,
in this paper we consider the so-called $t-J-U$ model, which is defined by
\begin{equation}
\label{eq:tJU}
H= -\sum_{\langle i,j \rangle,\sigma}t_{ij}(c^{\dagger}_{i,\sigma}
c_{j,\sigma} + {\rm {h.c.}})
+{\sum_{\langle i,j \rangle}}J_{ij}({\bf {S_{i} \cdot S_{j}}} 
-\frac{1}{4}n_{i}n_{j}) + U\sum_i n_{i,\uparrow}n_{i,\downarrow}~~,
\end{equation}
where the sites of a two-dimensional square lattice of size
$L_x \times L_y$ with periodic boundary
conditions are labeled by the indices $i$ and $j$, $t_{ij}$ and $J_{ij}$ 
are the hopping integrals and exchange couplings between sites $i$ and 
$j$, respectively, $c_{i,\sigma}$ is the annihilation operator for an 
electron of spin $\sigma$, $U$ is the on-site Hubbard repulsion energy, 
and $n_{i,\sigma}$ is the number operator for electrons at site $i$ with
spin $\sigma$.

The most familiar strong-coupling variant of the Hubbard model is the
$t-J$ model, and the physics of (square lattice) doped Mott insulators 
described by this model was reviewed by Dagotto \cite {elbio_rmp}. As 
emphasized by, {\it e.g.}, Anderson \cite {pwa_97}, the vital component 
of the $t-J$ Hamiltonian is the constraint of no double occupancy. That 
is, in the $t-J$ model one does not use the electron creation and 
annihilation operators of Eq.~(\ref{eq:tJU}), but rather one uses 
constrained creation and annihilation operators (for example, see the 
discussions in Ref.~\cite{elbio_rmp}). However, the above $t-J-U$ 
Hamiltonian can be used to accomplish this same mathematical projection 
by taking $U\rightarrow \infty$ --- this simplifying approach has been 
noted by a variety of researchers (see, {\it e.g.}, 
Refs.~\cite{lin,pethukov,kagan}), and will be used by us also.

In this paper we restrict $t_{ij}$ and $J_{ij}$ to be nonzero for near 
neighbours (NN) only.  Further, we allow the hopping integral in the 
$x$ direction, $t_x$, to be different than the hopping integral in the 
$y$ direction, $t_y$. We have also investigated the physics that
arises when $J_x$ is allowed to be different than 
$J_y$, but find that no qualitatively new physics arises as long as both
$J_x$ and $J_y$ remain nonzero. Thus, from now
on we set $J \equiv J_x = J_y$, and we analyze the resulting
hopping anisotropy problem in terms of 
\begin{equation}
\label{eq:anist}
t_x \equiv t~~,~~~~~r = \frac{t_y}{t}~~.
\end{equation}
By taking the limit $U/t \rightarrow \infty$ the only two 
(dimensionless) energy scales left in this problem are $J/t$ and $r$. 

The physics of the $t-J-U$ model for isotropic NN hopping ($r = 1$)
in the dilute electron density limit is clear: the antiferromagnetic 
exchange provides an attractive interaction between the electrons on 
neighbouring sites forming a singlet state; however, a bound state that 
is formed from this interaction must do so in a manner that excludes 
double occupancy, a result that must follow in the $U\rightarrow\infty$ 
limit.
As mentioned above, this problem for isotropic hopping and 
exchange has already been studied elsewhere \cite {lin,pethukov}. 
Here we shall analyze the bound state exchange threshold, to be denoted 
by $J_c/t$, as a function of hopping anisotropy $r$.

We consider the situation in which there are only two electrons present, 
in which case we can write for the singlet wavefunction of a two-electron 
system as\cite{lin,pethukov}
\begin{equation}
\label{eq:phi_real}
\psi = \sum_{i,j} \phi(i,j)c^{\dagger}_{i,\uparrow}
c^{\dagger}_{j,\downarrow}|0\rangle
\end{equation}
where $|0\rangle$ denotes the empty lattice. Any bound state can
be solved for by the direct solution of the Schr\"odinger equation.
Expressing $\phi(i,j)$ in terms of its Fourier transform,
the latter of which involves the centre of mass and relative
momenta, given by ${\bf Q} = {\bf k_{1}} + {\bf k_{2}}$ and ${\bf q}=
\frac{1}{2}({\bf k_{1}- k_{2}})$, with $\phi
({\bf {k_{1},k_{2}}})= \phi_{\bf {Q}}({\bf {q}})$, where 
\begin{equation}
\label{eq:phi_recip}
\phi({\bf {k_{1},k_{2}}}) = \frac{1}{N}\sum_{i,j}e^{-i{\bf k_1} \cdot 
{\bf r}_i - i {\bf k_2} \cdot {\bf r}_j}\phi(i,j)
\end{equation}
with $N=L_x L_y$, the resulting equation for the bound-state 
energy $E$ and its associated wave function $\phi_{\bf {Q}}({\bf {q}})$ is
\begin{equation}
\label{eq:Schro_eq}
\phi_{\bf Q}({\bf q}) = \frac{\frac{U}{N}\sum_{\bf k}
\phi_{\bf Q}({\bf k}) -\frac{1}{N}\sum_{\bf k}J({\bf q-k})
\phi_{\bf Q}({\bf k})}{E -\varepsilon(\frac{\bf Q}{2}+{\bf q}) 
-\varepsilon(\frac{\bf Q}{2}-{\bf q)}}~~.
\end{equation}
The dispersions $\varepsilon({\bf k})$ and $J({\bf k})$, the former of
which includes the hopping anisotropy, are given by
$\varepsilon({\bf k}) = - 2t ( \cos k_{x} + r\cos k_{y} )$ and
$J({\bf k}) = 2J ( \cos k_{x} + \cos k_{y} )$~.

The lowest energy bound states for the $t-J-U$ model has zero 
centre-of-mass momentum (${\bf Q}=0$). Then, one can solve 
Eq.~(\ref{eq:Schro_eq}) via the $3\times 3$ set of algebraic equations:
\begin{eqnarray}
\label{eq:3x3eqs}
C_{0} & = & UI_{0}C_{0} - 2JI_{x}C_{x} - 2JI_{y}C_{y} \\ \nonumber
C_{x} & = & UI_{x}C_{0} - 2JI_{xx}C_{x} - 2JI_{xy}C_{y} \\ \nonumber
C_{y} & = & UI_{y}C_{0} - 2JI_{xy}C_{x} - 2JI_{yy}C_{y} 
\end{eqnarray}
where
\begin{equation}
\label{eq:Cs}
C_{0}  =   \frac{1}{N}\sum_{{\bf {k}}}\phi({\bf {k}}) ~~~~
C_{x}  =  \frac{1}{N}\sum_{{\bf {k}}}\cos k_{x}\phi({\bf {k}}) ~~~~
C_{y}  =  \frac{1}{N}\sum_{{\bf {k}}}\cos k_{y}\phi({\bf {k}}) 
\end{equation}
and
\begin{eqnarray}
\label{eq:elipints}
I_{0}  = \frac{1}{N}\sum_{{\bf {q}}}\frac{1}{E + 4t(\cos q_{x} + 
r\cos q_{y})} 
&~~~~~~~~&
I_{x} = \frac{1}{N}\sum_{{\bf {q}}}\frac{\cos q_{x}}{E + 4t(\cos q_{x} + 
r\cos q_{y})}
\\ \nonumber
I_{y}  = \frac{1}{N}\sum_{{\bf {q}}}\frac{\cos q_{y}}{E + 4t(\cos q_{x} + 
r\cos q_{y})} 
&~~~~~~~~&
I_{xy}  = \frac{1}{N}\sum_{{\bf {q}}}\frac{\cos q_{x}\cos q_{y}}{E + 
4t(\cos q_{x} + r\cos q_{y})} \\  \nonumber
I_{xx}  = \frac{1}{N}\sum_{{\bf {q}}}\frac{\cos^{2} q_{x}}{E + 
4t(\cos q_{x} + r\cos q_{y})} 
&~~~~~~~~&
I_{yy}  = \frac{1}{N}\sum_{{\bf {q}}}\frac{\cos^{2} q_{y}}{E + 
4t(\cos q_{x} + r\cos q_{y})} \\ \nonumber
\end{eqnarray}

A methodology leading to the solution of such a set of equations is 
presented
in the papers of Lin\cite{lin} and Pethukov, {\it et al.}\cite{pethukov} 
However, in those papers the authors were considering a two-dimensional 
square plane with isotropic hopping, and thus were able to exploit 
$C_x = C_y$ owing to the result that the ground state wave function has 
s-wave symmetry. In our problem the Hamiltonian has a reduced symmetry,
leading to $C_x \ne C_y$ (for $r \not=1$), and thus we have employed
a different approach --- we will explain this method in detail in a
future publication \cite{sb_rjg}.

\begin{figure}
\begin{center}
\epsfig{file=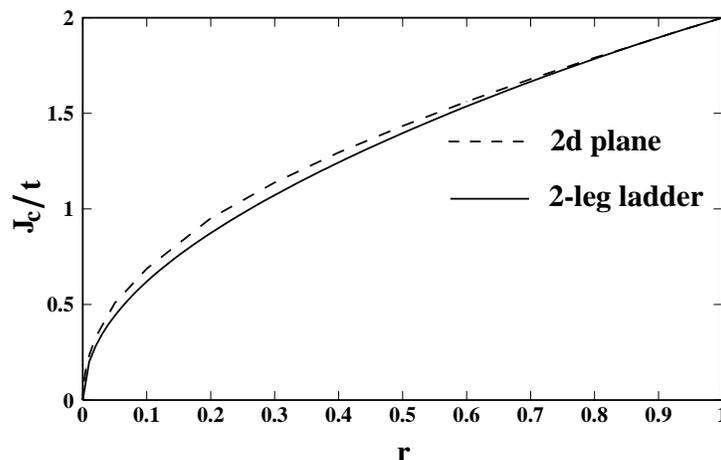,scale=0.75}
\caption{
The dashed line is a plot of Eq.~\protect (\ref{eq:Jc_2d}), and shows
how the threshold exchange for bound state formation, $J_c/t$, varies
as a function of hopping anisotropy, $r$, for an infinite two-dimensional
plane. The solid line is a plot of Eq.~\protect (\ref{eq:Jc_2leg}),
which is the same quantity for a two-leg ladder. Note the rapid decrease
to $J_c/t = 0$ for both systems as $r \rightarrow 0$.
}
\label{fig:Jcvsr}
\end{center}
\end{figure}

\begin{figure}
\begin{center}
\epsfig{file=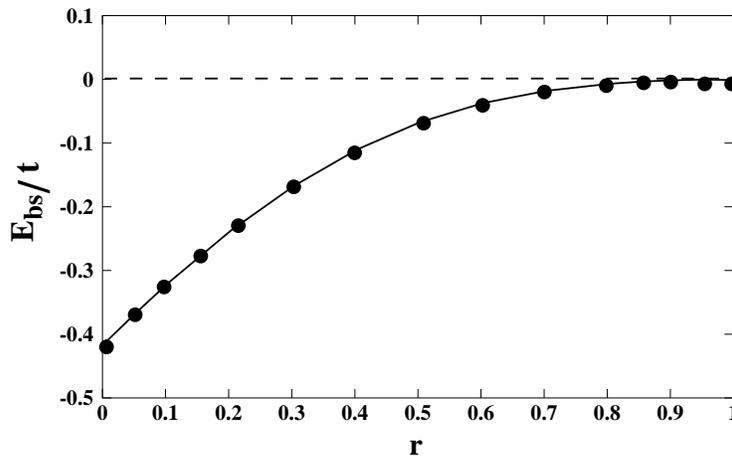,scale=0.75}
\caption{
The value of the bound state energy, $E_{bs}$, relative to the
bottom of the two-electron noninteracting band, which is equal to
$-4t(1+r)$, as a function of $r$ for $J/t=2$ for the infinite
two-dimensional plane. (Note that for this value of $J/t$ a bound state 
is found for all $r$.)
}
\label{fig:Ebvsr}
\end{center}
\end{figure}

We find that the resulting bound state is always symmetric under $\pi/2$
rotations, and reflections along the $x$ or $y$ directions; that is,
the ground state transforms under the identity representation of
the rectangular point group, and thus continuously interpolates with 
the s-wave bound state for $r=1$. The threshold $J_c/t$ as a function
of $r$ is given by the following expression:
\begin{equation}
\label{eq:Jc_2d}
\frac{J_c}{t} ~=~ -\frac{16\pi^{2}r^{5/2}}{g(r)} 
\left [f(r)-\sqrt{f(r)^{2}-
\frac{g(r)}{64\pi^{3}r^{3}}} \right ]
\end{equation}
\noindent where $f$ and $g$ are functions given by
\begin{eqnarray}
\label{eq:fg}
f(r)&=& \frac{1}{32}\frac{\left [4(r^{3}-r^2+r-1)\sin^{-1}
\sqrt{\frac{r}{r+1}}-
- 2\pi r^{3} + 4r^{5/2}+ 2\pi r^{2} +4\sqrt{r} \right ]}
{\pi^{2}r^{5/2}} \\ \nonumber
g(r)&=&2(r^2-1)\sin^{-1}\sqrt{\frac{r}{r+1}}-\pi r^{2} +2r^{3/2} 
+ 2\sqrt{r}
\end{eqnarray}
The resulting $J_c/t~vs.~r$ is plotted in Fig.~\ref{fig:Jcvsr}; from 
this plot we note (i) the $r=1$ limit of $J_c/t=2$ result is found, 
confirming the known isotropic result\cite{lin,pethukov}, and (ii) the
surprising result that as $r\rightarrow 0$, $J_c/t \rightarrow 0$.
This latter result is important --- it implies that one can form
bound states for arbitrarily small $J/t$ if one is able to create
a situation in which the hopping anisotropy is large enough. Further,
it implies that for a fixed $J/t$, the binding energy of two
electrons will become increasingly larger as this anisotropy is increased.
We have plotted this bound state energy result in 
Fig.~\ref{fig:Ebvsr}, for $J/t = 2$, from which
it is seen that one can obtain remarkably large binding energies. While
acknowledging our inability to relate this system directly to the 
high-T$_c$
systems, without further trepidation we evaluated the $r \rightarrow 0$ 
limit of the binding energy for $J/t=0.3$. We found a binding energy of
$0.0068 t$, and in units in which $t$ is roughly an eV, this implies
a binding energy of roughly 100 K, a provocative energy.

The question remains, what are the nature of the correlations in
such bound states? In an attempt to answer this question we have
solved the same problem as above not for an infinite two-dimensional
plane, but for a two-leg ladder. Technically, this involves nothing
more than using
\begin{equation}
\label{eq:2legqsums}
\sum_{\bf {q}} \rightarrow \frac{1}{2}\sum_{q_{y}=0,\pi}~~~
\frac{1}{2\pi}\int^{+\pi}_{-\pi}dq_{x}
\end{equation}
in the wave vector sums that appear in the evaluation of
the various (elliptic) integrals of 
Eqs.~(\ref{eq:Cs},\ref{eq:elipints}).
The resulting $J_c/t$ as a function of $r$ is given by
\begin{equation}
\label{eq:Jc_2leg}
\frac{J_c}{t}=
\frac{\left((1+r)^2+2\sqrt{r+r^2}-\sqrt{(1+r^2)^2+4r}\right )}
{r+1}
\end{equation}
and is also shown in Fig.~\ref{fig:Jcvsr}. 
Thus, we see that both the infinite
two-dimensional plane and the infinitely long two-leg ladder behave 
in a similar fashion. Further, the non analytic behaviour of
$J_c/t$ is quantitatively similar in the extreme hopping anisotropy 
limit: in the limit of $r\rightarrow 0$ the two-dimensional infinite 
plane has $J_c/t \sim (3\pi/4) \sqrt{r}$, while the two-leg ladder has 
$J_c/t \sim 2 \sqrt{r}$. 

\begin{figure}
\begin{center}
\epsfig{file=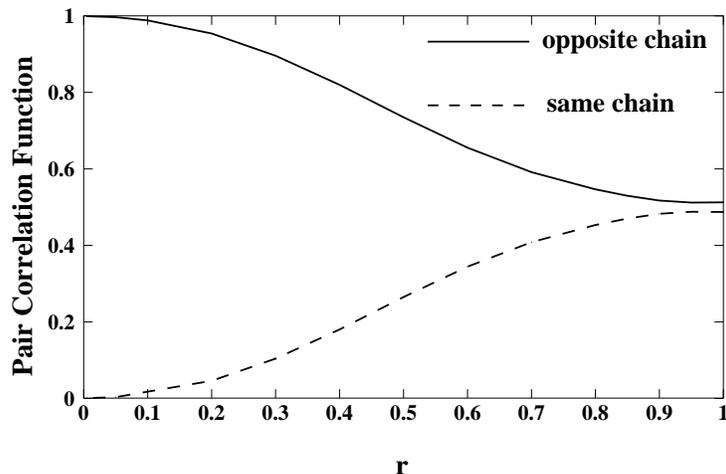,scale=0.75}
\caption{ 
The pair correlation for the bound state of the two-leg ladder was 
evaluated for all electron positions, and then summed to yield the 
probabilities that the electrons are on the same chain or on opposite 
chains. This quantity is shown as a function of hopping anisotropy, $r$.
Note that as $r \rightarrow 0$ the electrons are restricted to 
be on opposite chains only.
}
\label{fig:Pofrvsr}
\end{center}
\end{figure}

We have exploited this similarity by evaluating the pair correlation
function for the two-leg ladder, noting that the simpler geometry
of the two-leg ladder more readily allows for us to understand the
spatial character of the extreme hopping anisotropy bound states.
More specifically, we have calculated
the probability that the two electrons are on the same chain or
are on opposite chains. Our results as a function of $r$ are shown in
Fig.~\ref{fig:Pofrvsr}, from which it is seen that the character of the 
small $r$ bound
states is two electrons moving on opposite chains, with the pairing
resulting from a spin exchange interaction between the chains (when
the electrons are on neighbouring sites). 
There is a higher energy
bound state that corresponds to $J_c/t=2$, but in this bound state
the systems behaves like two one-dimensional chains, and the threshold
for this higher energy bound state is the same as for a one-dimensional
$t-J-U (U\rightarrow \infty)$ model, {\it viz.} $J_c/t=2$. Thus, the
lower energy and lower threshold bound states can only be achieved when
the electrons are moving along opposite chains. 

This situation is qualitatively similar to two electrons moving in 
neighbouring stripes that still experience an exchange coupling between 
the stripes due to the antiferromagnetically correlated domain between 
the stripes, but which have an effective hopping frequency that is much 
larger along the metallic rivers of charge than between such rivers.
In this very simplistic picture, it seems that stripe correlations
leading to parallel rivers of charge separated by antiferromagnetically
correlated domains can (at least) enhance bound state formation.

We have been able to complete the above work analytically because
of the relative simplicity of the zero density two-electron problem. 
Of course, the natural question is: Does this extrapolate to systems
that are doped away from half-filling? To address this question
we have considered the problem of two holes in an otherwise half-filled 
$12\times 2$ two-leg ladder for $J/t = 0.3$. For all $r$ the ground
state is ${\bf Q}=0$ and transforms under the identity representation
of the rectangular point group. The binding energy is again found to be 
substantially enhanced by the hopping anisotropy, a result
previously noted in Ref. \cite {weihong} --- our data is
listed below in Table~I. Lastly, we again evaluated the probability
that the holes existed on the same or opposite chains of the two-leg
ladder, and our results are qualitatively similar to those of
the two-electron problem that we have shown in Fig.~\ref{fig:Pofrvsr} 
(the two-hole correlations near $r=1$ have a greater probability of 
being on the opposite chains (about 0.62), but otherwise these two 
systems behave in very similar ways).

\begin{table}
\caption{The two hole binding energy, $E_{b}$, defined as  
$E_{b} = E_{2h}- 2E_{1h}+E_{0h}$, calculated for a 
$12\times 2$ two-leg  $t-J$ ladder with $J/t = 0.3$, as a 
function of anisotropy parameter $r$.}
\begin{center}
\begin{minipage}[t]{8cm}
\begin{tabular}{ccccc} 
 &  $r$  &    $E_b$    \\ \hline
 &  1.0  &    -0.1189  \\ 
 &  0.8  &    -0.1614  \\
 &  0.6  &    -0.2224  \\
 &  0.4  &    -0.2803  \\
 &  0.2  &    -0.3137   \\
 &  0.1  &    -0.3254   \\
\end{tabular}
\end{minipage}
\end{center}
\end{table}

Thus, the zero density two-electron problem
and the half filling two-hole problem behave in a very similar manner,
and are suggestive of the benefits of stripes in forming bound states.
Whether or not this physics extrapolates to pairing instabilities and,
{\it e.g.}, enhances the mean-field superconducting transition
temperatures, is being investigated presently.

\nopagebreak

\acknowledgements
We thank David Johnston and Massimiliano Capezzali for many helpful 
comments, and Andrew Callan-Jones and Fred Nastos for assistance. 
This work was supported in part by the NSERC of Canada.

\end{document}